\documentclass[12pt]{article}
\title{NEUTRON-ANTINEUTRON OSCILLATIONS IN THE TRAPPING BOX\\}
\author{B.O. Kerbikov\\ State Research
Center\\Institute of Theoretical and Experimental Physics, \\
Moscow, Russia}
 \date{}
  \newcommand{\be}{\begin{equation}}
\newcommand{\ee}{\end{equation}}
\def\la{\mathrel{\mathpalette\fun
<}} 
\def\fun#1#2{\lower3.6pt\vbox{\baselineskip0pt\lineskip.9pt
\ialign{$\mathsurround=0pt#1\hfil ##\hfil$\crcr#2\crcr\sim\crcr}}}

\begin{document}
\maketitle

\begin{abstract}
 We  have reexamined the problem of $n-\bar n$ oscillations for
 ultra-cold neutrons (UCN) confined within a trap. We have shown
 that the growth of the $\bar n$ component with time is to a decent accuracy  given by
 $P(\bar n)= \varepsilon^2_{n\bar n} t_Lt,$ where $\varepsilon_{n\bar n}$ is the mixing parameter,
 $t_L\sim 1 $ sec in the
 neutron propagation time between subsequent collisions with the trap walls. Possible corrections  to this
 law and open questions are discussed.
\end{abstract}

\section{Introduction}

During several decades the problem of nucleon instability is a
subject of intense and diversified theoretical and experimental
studies. An interesting facet of this fundamental problem is a
hypothetical process of neutron-antineutron oscillations \cite{1}.
Such oscillations have been thoroughly discussed in free-space
regime and inside nuclei -see e.g. \cite{1}-\cite{5} and
references therein. The third and very interesting possibility to
search for $n-\bar n$ oscillations is to use ultra-cold neutrons
(UCN) confined in  a trap. This subject was discussed by several
authors \cite{2,3,6} but in contrast to the first two regimes the
picture of $n-\bar n$ oscillations of UCN remains rather obscure.
On the other hand several experiments of this kind are in
preparation now. Therefore it   is appropriate to address this
subject again. We shall follow two complementary lines of
arguments. The first one is based on  simple qualitative estimates
while the second makes use of the time evolution equation focusing
on the interaction of the two-component $n-\bar n$ system with the
walls of the trap. Both approaches lead to a conclusion that  the
$\bar n$ component grows (approximately) linearly with observation
time contrary to quadratic time dependence in the free-space
regime.

\section{Relevant Parameters}

We start by introducing  a set of  definitions and parameters. Our
treatment will be somewhat schematic in the  sense that  we do not
consider any specific geometry of the trap,  concrete UCN
spectrum, and a  variety of the trap  materials. All these points
can be easily accounted for as soon  as one sticks to a given
experimental setup.

First we remind that neutrons with energy $E<10^{-7}$eV= $100$neV
are called ultra-cold.  A useful relation connecting the neutron
velocity $v$ in cm/sec and $E$ in neV reads \be v({\rm cm/sec})
\simeq  10^2\{E({\rm neV})/5.22\}^{1/2}.\label{1}\ee
 In particular the velocity corresponding to $E=100$ neV is
 $v\simeq4.4\cdot 10^2$cm/sec.

 A less formal definition of UCN involves a notion of the  real
 part of the optical potential  corresponding to the trap
 material. Namely,  neutrons  with energies less than the  height
 of this potential are called ultra-cold. The two definitions are
 essentially equivalent since for most materials the optical
 potential is of the order of 100 neV (see  below).

 Our main interest concerns strongly absorptive interaction of
 antineutron component with the trap wall. Therefore very weak
 absorption of    neutrons on the wall will be neglected.
 Interesting by itself this problem is out of the scope  of the
 present work. According to the second definition of UCN
 (i.e. $E<U_{nA}$) they undergo   complete   reflection  from the
 trap walls and may be stored for about $10^3$ sec ($\beta$ -
 decay time) as was first point out by Ya.B.Zeldovich \cite{7}.
 For each  material the limiting neutron velocity is given by  (\ref{1})
 with $E$ substituted by $U_{nA}$.

 To be concrete we consider neutrons with energy $E=80$ neV which
 according to (\ref{1})
corresponds to  $v=3.9\cdot 10^2$cm/sec. Such neutrons have
momenta $k\simeq 12.3$ eV, de Broglie wave lengths $\lambda\simeq
10^{-5} cm.$ As for the wall material we take $C^{12}$ with the
density $\rho=2.25g/cm^3$, or $N\simeq 1.13\cdot 10^{-16}
fm^{-3}$. The coherent $nC^{12}$ scattering length is $a_n=6.65 fm
$ \cite{8} (the imaginary part of the scattering length is at
least 3 orders of magnitude smaller and is  ignored as was already
stated). The corresponding $nC^{12}$ optical potential reads
\be
U_n=\frac{2\pi}{m} Na_n\simeq 195 neV,\label{2}\ee  with $m$ being
the neutron mass.  Limiting velocity corresponding to $E=U_n$ is
$v_n^{(0)}\simeq 6.1\cdot 10^2$ cm/sec. Thus neutrons under
consideration with $v_n=3.9\cdot 10^2$ cm/sec are certainly
ultra-cold with respect to $C^{12}$ trap walls.

 The experimental data on antineutron-nuclei scattering lengths at
 ultra-low energies are absent. Only some indirect information may
 be gained from the level shifts of antiprotonic atoms. The energy
 behavior of the $\bar nA$ annihilation cross-section is governed
 by the  well-known
" $1/v$" law
 \be
 \sigma_a=-4\pi\frac{Ima_{\bar n A}}{k}.\label{3}
\ee

Several fits to the $\bar n A$ scattering lengths have been
proposed in the literature. We consider as most reliable that of
Ref. \cite{9} based on internuclear  cascade model. Even within
this particular model one finds several solutions for $\bar
nC^{12}$ scattering length. Therefore the one we have chosen for
our analysis may be called "motivated" by Ref. \cite{9} and reads
\be
a_{\bar n}=(3-i1)fm.\label{4}\ee

Then according to (\ref{2}) the  $\bar nC^{12}$ optical potential
is equal to
\be
V_{\bar n} =U_{\bar n} -iW_{\bar n} \simeq (90-i30) {\rm
neV}.\label{5}\ee

The  limiting $\bar n$ velocity corresponding to $U_{\bar n} $
from (\ref{5}) is according to (\ref{1}) $v^{(0)}_{\bar n}\simeq
4.15\cdot 10^2${cm}/{sec}. Now our choice $E=80 neV < U_{\bar n}
<U_n$ is clear since we want to deal with "ultra-cold" $\bar n$ as
well. The case $U_{\bar n}< E< U_n$ will be considered in the next
publication. Needless to say that due to strong $\bar n$
absorption the condition $E_{\bar n} < U_{\bar n}$ in no way
provides complete reflection of $\bar n$ from the wall -- see
below.

Next we remind that the lower limit on $n-\bar n$ oscillation time
$\tau_{n\bar n}$  has been  obtained from experimental study of
$n-\bar n$ transitions in free space and inside nuclei -- see
\cite{10,4,11,9} and references therein. For our purposes it is
enough to  keep in mind a  crude value
\be
\tau_{n\bar n} >10^8 {\rm sec}.\label{6}\ee

Correspondingly the value of the mixing parameter is
\be
\varepsilon_{n\bar n} =\frac{1}{\tau_{n\bar n}}<10^{-23} {\rm
eV}.\label{7}\ee
 The dynamical meaning of $\varepsilon_{n\bar n} $
will be clear from the evolution equation which will be presented
below. To give some perception of the value of $\varepsilon_{n\bar
n} $ we may say that if one considers neutron confined in a
one-dimensional 5 meters long box then the level splitting will be
just $10^{-23}$eV.

Finally we introduce the parameter $t_L\sim 1 sec$ -- the time
which neutrons need to cross the trapping box, or the time between
two subsequent collisions with the walls. We also remind that both
$\beta$-decay time and the UCN storage time are of the order of
$10^3$ sec.

\section{Hitting the Trap Wall: Simple Estimates}

Now with relevant parameters at hands we can  analyze what happens
with  neutrons  and possible antineutrons  admixture when they
interacts with the wall of the trap. In this section we present
simple estimates.

Treatment based on time evolution equation will be postponed till
 the next section.
A third approach based on the wave packets formulation will be
only touched on  in  present paper and discussed in detail in the
next publication.

First we consider the collision of neutrons with the wall. As
compared to antineutrons this problem is much simpler due to the
lack of absorption (see remarks on the previous pages).

In our illustrative example a neutron with  $v=3.9\cdot 10^2 $
cm/sec is hitting the $C^{12}$ wall of the trap. Such a velocity
is well below the limiting $C^{12}$ value $v^{(0)}_n\simeq
6.1\cdot 10^2$ cm/sec and hence neutron undergoes a complete
reflection, $R=1$. Being absolutely correct, the statement that
the reflection coefficient $R=1$ does not constitute the whole
story. First, even at
 $v<v^{(0)}_{\bar n} $
 the tail of the neutron
  wave function  penetrates inside the wall. On general
 grounds the penetration  depth  is  $l_w (n)\sim
\lambda\simeq 10^{-5}$ cm,  with $\lambda$ being de Broglie wave
length. Second, collision with the wall is not an instantaneous
act but is characterized by certain collision time. The rigorous
derivation of this time should be based on the wave packets
formalism \cite{12}-\cite{14}. However simple estimates presented
below yield the same results.

Inside the wall the neutron wave function has the form
\be
\psi_{\bar n} (x)\propto\exp\{-\kappa_n x\},~~\kappa_n= \sqrt{2m
(U_{n} - E)}. \label{8}\ee

From {\ref{8}) it is natural to identify

\be
l_w(n)\sim 1/\kappa_n\simeq 0.14\cdot 10^{-5} {\rm cm},
\label{9}\ee which is few times less than the naive expectation
$l_w(n)\sim \lambda= 10^{-5}$ cm. Collision time may be estimated
as\footnote{ Strictly speaking neutron velocity inside the wall is
different from $v$.}
\be
\tau_{coll} (n) \sim \frac{2l_w (n)}{v} \simeq 0.7\cdot
10^{-8}{\rm sec}.\label{10}\ee

This result is in perfect agreement with what is predicted from
collision theory formulated in terms of the wave packets. Namely,
if one describes the incident neutron by a wave packet
\be
\psi_n(x,t) =\sqrt{\frac{a}{\pi}} \exp (ikx -\frac{k^2t}{2m})
\frac{\sin [(x-vt)/a]}{x-vt},\label{11} \ee where $a$ is its width
in $x$-space, then
\be
\tau_{coll} = [E(U_0-E)]^{-1/2}\simeq 0.7\cdot 10^{-8}{\rm
sec}\label{12}\ee for $E=80$ neV, $U_0=195$ neV. At this point we
note that $\tau_{coll} v=2l_w\ll a =\frac{\lambda}{\pi} \left(
\frac{\Delta\lambda}{\lambda}\right)^{-1}$. We shall return to
this remark in connection with possible decoherence of $n$ and
$\bar n$ due to the difference in their collision times.

Now we turn to $\bar n$ with the same velocity $v=3.9\cdot 10^2$
cm/sec hitting the $C^{12}$ wall.

Due to absorption (annihilation) the $\bar n$ wave function inside
the wall has the form
\be
\psi_{\bar n} (x) \propto e^{ik_{\bar n} x-\kappa_{\bar
n}x},\label{13}\ee
\be
(ik_{\bar  n}-\kappa_{\bar n})^2= 2m(U_{\bar n} -iW_{\bar n} -
E),\label{14} \ee where $E$ is the energy of the incident $\bar
n$. Eqs. (\ref{13}-\ref{14}) yield
\be
l_w(\bar n) \simeq \frac{1}{\kappa_{\bar n}}= m^{-1/2}\left
\{U_{\bar n}-E+[(U_{\bar n}-E)^2+W^2]^{1/2} \right\}^{-1/2} \simeq
0.32\cdot 10^{-5}{\rm cm}.\label{15}\ee Then $\bar n$ collision
time is
\be
\tau_{coll} (\bar n)\sim \frac{2l_w(\bar n)}{v} \simeq 1.6\cdot
10^{-8} {\rm sec},\label{16}\ee which  is about  2  times larger
than the neutron collision time given by (\ref{10})\footnote{Wave
packets formalism, as was shown by V.A.Lensky, leads to somewhat
smaller value -- see our next publication.}.

The crucial parameter which determines the fate of $\bar n$
hitting the wall is the ratio of the collision time (\ref{16}) to
the absorption (annihilation) time. The later quantity is velocity
independent by virtue of the "$1/v$" law (\ref{3}) and is
expressed through the $\bar n$ mean free  path $\Lambda$ according
to
\be
\tau_{abs} (\bar n) \sim \frac{\Lambda}{v} \simeq \frac{m}{4\pi
N|Ima_{\bar n A}|}\simeq 1.1\cdot 10^{-8} {\rm sec}\label{17}\ee
for the $C^{12}$ trap wall. Thus \be \tau_{coll} (\bar
n)/\tau_{abs} (\bar n)>1, \label{18}\ee which implies the collapse
of the possible $\bar n$ component on the wall.

Already at this point it is clear that this in turn leads to the
time dependence of the probability to find $\bar n$ component
announced in the Abstract, namely $ P(\bar n) =\varepsilon^2
t_Lt$. A rigorous derivation of this equation is given in the next
section.

Still one may argue that the  above estimates should be taken with
caution and certain fraction of $\bar n$ may be still reflected
from the wall. Then Eqs.(\ref{10}) and (\ref{16}) enable to
estimate the  splitting between the centers of the $n$ and $\bar
n$ wave packets (see (\ref{11})) after the reflection. One has
\be
\delta x\simeq v (\tau_{coll} (\bar n)-\tau_{coll} (n)) \simeq
0.35\cdot 10^{-5} cm \sim \lambda\ll a.\label{19}\ee Whether this
retardation influences the $n-\bar n$ mixing in free space between
collisions with the trap walls will be discussed elsewhere.

The main point to be improved on in the above estimates is
mentioned in the footnote to Eq.(\ref{10}). Certain guidance in
this direction may be found in \cite{15}.

Finally we note that the treatment presented above seems
physically more transparent  than formal calculations of the
reflection coefficient from the complex potential.

\section{Hitting the Trap Wall: Time-Dependent Approach}

As a "warming up" exercise we consider $n-\bar n$ oscillations in
a free space with  $\beta$- decay neglected. This is a standard
two-level problem treated in any serious textbook  on Quantum
Mechanics. The phenomenological Hamiltonian is a $2\times 2$
matrix in the  basis of the two-component $n-\bar n$ wave function
\be
H=E_i\delta_{ij} +\varepsilon \sigma_x,\label{20}\ee with
$i,j=n,\bar n$. The evolution equation reads \be
i\frac{\partial}{\partial t} \left( \begin{array}{l} \psi_n\\
\psi_{\bar n}\end{array} \right)= \left( \begin{array}{ll}
E_n&\varepsilon\\\varepsilon &E_{\bar n}\end{array} \right) \left(
\begin{array}{l} \psi_n\\ \psi_{\bar n}\end{array}
\right).\label{21}\ee Assuming that $\psi_n(t=0)=1, ~~\psi_{\bar
n} (t=0)=0$, and diagonalizing the Hamiltonian (\ref{20}) one
arrives at the following expression for the probability of finding
$\bar n$ at a time $t$ \cite{2}-\cite{6}
\be
\psi_{\bar n} (t) |^2 =\frac{4\varepsilon^2}{\omega^2
+4\varepsilon^2} \sin^2 (\frac12 \sqrt{\omega^2+4\varepsilon^2}
t), \label{22}\ee where $\omega=(E_{\bar n} -E_n)$. In free space
the difference between $E_{\bar n}$ and $E_n$ may be due to the
Earth magnetic field. In this case \be\omega=2\mu_nB\simeq 6\cdot
10^{-12} {\rm eV}. \label{23}\ee Without magnetic field, i.e. at
$\omega=0,$ and at $t\ll \tau_{n\bar n}\simeq  10^8$ sec one has
\be |\psi_{\bar n} (t)|^2\simeq \varepsilon^2_{n\bar n}
t^2,\label{24}\ee while with the Earth magnetic field
Eq.(\ref{24}) is valid only at extremely short times, $t\ll (\mu
B)^{-1}\simeq 2\cdot 10^{-4}$ sec, while at large times \be
|\psi_{\bar n} (t)|^2\simeq\frac{4\varepsilon^2}{\omega^2} \sin^2
t/\tau_B\simeq 10^{-23}\sin^2 t/\tau_B, \label{25}\ee
 where $\tau_B=(\mu B)^{-1} \simeq 2\cdot 10^{-4}$ sec.

 The use of (\ref{22}) to test fundamental symmetries is discussed
 in \cite{5}.

 Next we consider the general Hamiltonian of the $n -\bar n$
 system inside the wall with annihilation and $\beta$-decay
 included. The problem is reminiscent of  strangeness oscillations
 in $K\bar K$ system. With annihilation and $\beta$-decay included
 the Hamiltonian (\ref{20}) is substituted by \be
 H=\left( \begin{array}{ll}
 E_n-i\frac{\Gamma_\beta}{2}&\varepsilon\\
 \varepsilon& E_{\bar
 n}-i\frac{\Gamma_a}{2}-i\frac{\Gamma_\beta}{2}\end{array}\right),\label{26}\ee
 where $\Gamma^{-1}_\beta\sim 10^3$sec, $\Gamma_a\simeq 2W_n\simeq
 60$ neV for $C^{12}$.

 In arriving  to (\ref{22}) diagonalization of the Hamiltonian
 (\ref{20}) has been done exactly. Performing similar procedure
 with (\ref{26}) use can be made of a small parameter
 $4\varepsilon^2 \ll|H_{11}-H_{22}|^2$. Indeed, inside the wall
 effective fields acting on $n$ and $\bar n$ differ by  tens of
 neV (see (\ref{2}) and (\ref{5})) while $\varepsilon\sim 10^{-14}$
 neV. Expanding $\{(H_{11}-H_{22})^2+4\varepsilon^2\}^{1/2}$ with
 respect to this small parameter one finds the two eigenvalues of
 the Hamiltonian (\ref{26})
 \be
 \mu_1\simeq E'_n-i\frac{\Gamma_\beta}{2}
 -i\frac{\Gamma_\varepsilon}{2},\label{27}\ee

 \be
 \mu_2\simeq E'_{\bar n}-i\frac{\Gamma_a}{2}-i\frac{\Gamma_\beta}{2}
 +i\frac{\Gamma_\varepsilon}{2}.\label{28}\ee
Here $E'_n=E_n-E_\varepsilon,~~ E'_{\bar n}= E_{\bar
n}+E_{\varepsilon},$ and \be
E_\varepsilon+i\frac{\Gamma_\varepsilon}{2}
=\frac{\varepsilon^2}{E_{\bar n}-
E_n-i\frac{\Gamma_a}{2}}.\label{29}\ee The "wrong" sign of the
last term in (\ref{28}) is an artifact of the square root
expansion, but this is physically irrelevant since
$\Gamma_\varepsilon\ll\Gamma_\beta\ll
\Gamma_a(\Gamma_\varepsilon\sim  10^{-39} {\rm
eV},~~\Gamma_\beta\sim 10^{-18} {\rm eV},~~\Gamma_a\sim 10^{-7}
{\rm eV}).$

In terms of eigenvalues $\mu_1$ and $\mu_2$ the general solution
of the two-component evolution equation has the form \cite{16}\be
\psi(t) = \left( \begin{array}{l}\psi_n(t)\\\psi_{\bar
n}(t)\end{array}\right) = \left(
\frac{H-\mu_2}{\mu_1-\mu_2}e^{-i\mu_1 t}
+\frac{H-\mu_1}{\mu_2-\mu_1}e^{-i\mu_2 t}\right)
\psi(0).\label{30}\ee

Again we start with a solution corresponding to initial conditions
$\psi_n(t=0)=1,~~ \psi_{\bar n} (t=0) =0.$  Then from
(\ref{26})-(\ref{30}) one gets \be |\psi_{\bar n}
(t)|^2=\frac{\varepsilon^2}{\omega^2+\frac{\Gamma^2_a}{4}}e^{-(\Gamma_\beta+\Gamma_\varepsilon)t}\left\{
1+e^{-\Gamma'_at}-2e^{-\frac{\Gamma'_a}{2}t}\cos \omega
t\right\},\label{31}\ee where $\omega=(E_{\bar n}-E_n),~~
\Gamma'_a=\Gamma_a-2\Gamma_\varepsilon$, and $\Gamma_\varepsilon$
is defined by (\ref{29}). Since $\Gamma_\varepsilon
\ll\Gamma_\beta$ and $\Gamma_\varepsilon \ll\Gamma_a$, one can
rewrite (\ref{31}) in a simpler form without noticeable lost of
accuracy, namely \be |\psi_{\bar n}
(t)|^2=\frac{\varepsilon^2}{\omega^2+\frac{\Gamma^2_a}{4}}e^{-\Gamma_\beta
t}\left\{ 1+e^{-\Gamma_at}-2e^{-\frac{\Gamma_a}{2}t}\cos \omega
t\right\}\label{32}\ee

This equation resembles that giving the probability to find $\bar
K^0$ in initially pure $K^0$ beam. Notice that instead of the
overall factor $1/4$ for $K^0-\bar K^0$ system we find in
(\ref{32}) an extremely small factor
$\varepsilon^2(\omega^2+\Gamma^2_a/4)^{-1}\sim 10^{-32}(!)$
reflecting the fact that mixing is very small as compared to the
complex splitting of $n$ and $\bar n$ eigenvalues in the medium.

Consider (\ref{32}) at $t= \tau_{coll} (\bar n) \simeq 1.6 \cdot
10^{-8}$ sec (see (\ref{16})). Then $\Gamma_\beta\tau_{coll} (\bar
n) \sim 10^{-11},~~ \Gamma_a\tau_{coll} (\bar n) \simeq 1.5, ~~
\cos \omega \tau_{coll} (\bar n) \simeq 0.7$, and  (\ref{32})
yields \be |\psi_{\bar n} (\tau_{coll} (\bar n))|^2\la 10^{-32}.
\label{33}\ee

Physically this means that if a pure  $n$ beam collides with a
wall made of $C^{12}$ the tiny admixture of  $\bar n$  which would
have emerged during the collision time is completely damped by
annihilation and $n-\bar n$ energy splitting.

The free-space regime (\ref{24}) is "hidden" in (\ref{32}) at the
limit of very short times, $ t\ll 1/\Gamma_a\sim 10^{-8}$ sec.
Then \be |\psi_{\bar n}
(t)|^2\simeq\frac{\varepsilon^2}{\omega^2+\frac{\Gamma^2_a}{4}}\sin^2\frac{\omega
t}{2}\simeq \frac{\varepsilon^2 t^2}{1+\Gamma_a^2/4\omega^2}
\simeq 0.9 \varepsilon^2 t^2,\label{34} \ee where at the last step
use has been made of the values of $\Gamma_a$ and $\omega$ for
$C^{12}$.

Next consider (\ref{30}) at initial conditions which are closer to
real experimental situation. Namely, suppose that UCN beam
collides with the wall after crossing the trap. The Earth magnetic
field is assumed to be shielded, so that the free-space equation
(\ref{24}) is valid inside the trap. Then initial conditions in
(\ref{30}) read \be \psi_{\bar n}(t=0) =\varepsilon t_L,~~ \psi_n
(t=0) =\sqrt{1-\varepsilon^2 t^2_L},\label{35}\ee where $t_L\simeq
1$ sec. Then at $t=\tau_{coll} (\bar n) \simeq 1.6\cdot 10^{-8}$
sec, i.e. just after the collision with the wall, one gets\be
|\psi_{\bar n} (\tau_{coll})|^2\simeq \varepsilon^2 t^2_L
e^{-(\Gamma_a+\Gamma_\beta)\tau_{coll}}\left[
1+0\left(1/t_L\Gamma_a\right)\right].\label{36}\ee This result is
again physically transparent. During the collision the $\bar n$
component is depleted by annihilation, while antineutrons "newly
born" inside the wall are damped according to our previous result
(\ref{32}).

Now $\tau_{coll} (\bar n) \simeq 1.6\cdot 10^{-8}$ sec,
$\Gamma_a\simeq 60$ NeV, so that $\exp
(-\Gamma_a\tau_{coll})\simeq 0.2$. This means that only 1 per
$\simeq 5\bar n$ survives after the collision.

This result  is in line with estimates  presented in Section 3 but
here we are on somewhat more qualitative footing.

If one considers the fraction 1/5 as a small parameter, then the
probability of antineutron detection at time $t$ will be
\be
|\psi_{\bar n} (t)|^2=\varepsilon^2 t_Lt,\label{37}\ee instead of
$\varepsilon^2 t^2$ free-space law (\ref{24}). Indeed, the
probability of $n-\bar n$ transition between the two subsequent
collisions with the walls is $\varepsilon^2 t^2_L$, while the
number of collisions during the observation time is
$\left(t/t_L\right)$. The extrapolation between the laws
(\ref{24}) and (\ref{37}) will be discussed in the next
publication.

\section{Acknowledgements}

The author is deeply grateful to Yu.A.Kamyshkov for grabbing him
into the subject and presenting a clear introduction. Discussion
of the oscillation problem with M.I.Vysotsky was very
enlightening. Useful remarks and information was gained from
V.A.Lensky, L.A.Kondratyuk,  A.E.Kudryavtsev and the members of
the ITEP seminar.

Special thanks for essential financial support are to V.A.Novikov,
L.B.Okun and grants RFFI 00-02017836 and 00-15-96786.

 \end{document}